\documentclass[12pt, a4paper]{article}
\usepackage[english]{babel}
\usepackage[utf8x]{inputenc}
\usepackage{amsmath}
\usepackage{graphicx}
\usepackage{subcaption} 
\usepackage{lscape}
\usepackage{tablefootnote}
\usepackage{tikz}
\usetikzlibrary{positioning, calc, shapes.geometric, shapes.multipart,
shapes, arrows.meta, arrows,
decorations.markings, external, trees}

\tikzstyle{Arrow} = [
thick,
decoration={
markings,
mark=at position 1 with {
\arrow[thick]{latex}
}
},
shorten >= 3pt, preaction = {decorate}
]

\usepackage{lmodern}
\usepackage{textcomp}
\usepackage[colorinlistoftodos]{todonotes}

\usepackage[mathscr]{euscript}
\usepackage[para,online,flushleft]{threeparttable}
\newcommand\independent{\protect\mathpalette{\protect\independenT}{\perp}}
\def\independenT#1#2{\mathrel{\rlap{$#1#2$}\mkern2mu{#1#2}}}

\usepackage{bm}
\usepackage{amsthm}
\usepackage{tikz}
\usetikzlibrary{arrows,automata,positioning}
\theoremstyle{definition}

\usepackage{booktabs,array}

\usepackage[plain,noend]{algorithm2e}

\usepackage{amsmath}
\usepackage{graphicx}

\usepackage{amssymb}

\newcommand{\bigCI}{\mathrel{\text{\scalebox{1.07}{$\perp\mkern-10mu\perp$}}}}
\newcommand{\rowgroup}[1]{\hspace{-1em}#1}

\usepackage[mathscr]{euscript}

\usepackage{titlesec}
\titleformat{\section}[block]
  {\center}
  {\thesection .}
  {1em}
  {\MakeUppercase}
\titleformat{\subsection}[hang]
  {\center \it}
  {\thesubsection}
  {1em}
  {}
\usepackage{authblk}
\author[1,2]{Anders Huitfeldt (*)}
\author[3,4]{Sonja A. Swanson}
\author[4,5]{Mats J. Stensrud}
\author[4,6]{Etsuji Suzuki}
\affil[1]{Norwegian Institute of Public Health}
\affil[2]{PharmacoEpidemiology and Drug Safety Research Group, Department of Pharmacy, and PharmaTox Strategic Initiative, Faculty of Mathematics and Natural Sciences, University of Oslo}
\affil[3]{Department of Epidemiology, Erasmus MC}
\affil[4]{Department of Epidemiology, Harvard T.H. Chan School of Public Health} 
\affil[5]{Department of Biostatistics, University of Oslo}
\affil[6]{Department of Epidemiology, Graduate School of Medicine, Dentistry and Pharmaceutical Sciences, Okayama University}

\begin{document}

\title{Effect heterogeneity and variable selection for standardizing causal effects to a target population}
\maketitle

\begin{abstract}
    
The participants in randomized trials and other studies used for causal inference are often not representative of the populations seen by clinical decision-makers. To account for differences between populations, researchers may consider standardizing results to a target population. We discuss several different types of homogeneity conditions that are relevant for standardization: Homogeneity of effect measures, homogeneity of counterfactual outcome state transition parameters, and homogeneity of counterfactual distributions. Each of these conditions can be used to show that a particular standardization procedure will result in an unbiased estimate of the effect in the target population, given assumptions about the relevant scientific context. We compare and contrast the homogeneity conditions, in particular their implications for selection of covariates for standardization and their implications for how to compute the standardized causal effect in the target population. While some of the recently developed counterfactual approaches to generalizability rely upon homogeneity conditions that avoid many of the problems associated with traditional approaches, they often require adjustment for a large (and possibly unfeasible) set of covariates.
\end{abstract}
\pagebreak
\section{Background}

The participants in randomized trials and other studies used for causal inference are often not representative of the populations seen by clinical decision-makers \cite{weiss_generalizing_2019-1}. Several statistical methods have been proposed to standardize estimates of a causal effect to the distribution of baseline covariates in a clinically relevant target population, in order to account for differences between populations. However, less attention has been given to how an investigator should reason about which covariates need to be standardized over. The choice of variables is important not only for the standardization procedure, but also for determining which personal characteristics of the participants must be considered when reasoning qualitatively about how representative a study is, relative to the intended target population. In this paper, we discuss different ways to select such covariates, and show that this problem is closely related to how one chooses to operationalize effect homogeneity between populations. 

For all examples, we will consider the effect of a binary treatment $A$ (for example, a pharmaceutical; 1 = treated, 0 = not treated) on a binary outcome $Y$ (for example, a side effect; 1 = occurred, 0 = did not occur). Counterfactuals will be denoted using superscripts. We will let $V$ denote a set of baseline covariates which are potential effect modifiers (for example: gender, nationality, etc). In some examples, we will consider an individual binary potential effect modifier (an element of $V$), which we will call $W$. We will consider two separate populations: The study population ($P=s$), in which we have valid evidence for the causal effect of the treatment; and the target population ($P=t$) in which we either have no data on the exposure or outcomes variables, or only have observational data and are unable to rule out confounding. We will consider membership in the study population to be defined at baseline, and focus on issues that arise due to non-random selection into the study. We note that while it is certainly possible that there is selection out of the study post-baseline, this is better considered as a form of selection bias \cite{hernan_structural_2004, suzuki_errors_2016} related to internal as opposed to external validity. We will focus primarily on binary outcomes, but note that most methods and concepts discussed in this paper (except counterfactual outcome state transition (COST) parameters) extend readily to continuous and time-to-event outcomes. We consider several measures of causal effect including the risk difference (RD), the risk ratio (RR), the survival ratio (SR) (which can be understood as the RR where the coding of the outcome variable is reversed), and the odds ratio (OR). These effect measures may be defined in a specific population or subgroup, which we denote using subscript as needed. For instance, $\text{RD}_t$ is the RD in population $t$.

Epidemiologists and clinical scientists have traditionally defined effect homogeneity in terms of a specific effect measure. For example, one may consider effect homogeneity as the absence of effect modification on the RR, RD or OR scale. These definitions of effect homogeneity are associated with several established conceptual and practical shortcomings, including lack of biological interpretation, baseline risk dependence, zero-bounds, prediction outside the range of valid probabilities, non-collapsibility and asymmetry \cite{huitfeldt_choice_2018}. There have also been several recent methodological developments in defining effect homogeneity based on counterfactual distributions rather than specific measures of effect \cite{dahabreh_extending_2019}. These approaches consider the outcome under the active treatment separately from the outcome under the control condition. VanderWeele described this as ``effect modification in distribution'' \cite{vanderweele_confounding_2012}, to contrast with the traditional approach, which was termed ``effect modification in measure''. 

Both these types of effect homogeneity may occur between two subgroups which are both in the study population (for example: between men in the study population and women in the study population), or between one subgroup which is in the study population and another subgroup which is in the target population (for example: between men in the study population and men in the target population). Homogeneity between two groups that are both in the study population is often invoked in meta-analysis and model specification. Homogeneity between one group that is in the study population and another that is in the target population is necessary in settings where the goal is to extrapolate the findings to settings outside of the observed data ("generalizability" or "transportability"). Conceptually, these types of homogeneity are closely related, and differ primarily in that the former is testable from the observed data, whereas the latter is not. 

The paper is organized as follows. First, we consider approaches to generalizability that are based on conditional homogeneity of standard effect measures (such as RR and RD) between the study population and the target population. We then describe the recently introduced COST parameters \cite{huitfeldt_choice_2018}, and show how this framework can be used to overcome some of the shortcomings of traditional effect measures. Finally, we review approaches to generalizability based on conditional homogeneity of individual counterfactual distributions, with a particular emphasis on methods based on inverse probability weighted estimators, and methods based on causal diagrams. As we introduce each approach, we repeatedly refer to two tables throughout the text: Table 1 shows an overview of different ways an investigator can operationalize effect homogeneity; Table 2 shows five different approaches to standardization which rely on different homogeneity conditions. Proofs of the standardization formulas in Table 2 are shown in appendix 1.

\begin{table}
\caption{Definitions of conditional effect homogeneity between study population and target population}
\centering
\begin{tabular}{p{0.4\linewidth}p{0.5\linewidth}}
\toprule
Homogeneity condition & Definition\\

\midrule
\rowgroup{Effect Homogeneity in Measure} \\
On the risk difference scale & $\text{RD}_{s,v} = \text{RD}_{t,v}$   \\
On the risk ratio scale &  $\text{RR}_{s,v} = \text{RR}_{t,v}$    \\
On the survival ratio scale &  $\text{SR}_{s,v} = \text{SR}_{t,v}$    \\
On the odds ratio scale & $\text{OR}_{s,v} = \text{OR}_{t,v}$    \\
\midrule
\rowgroup{Homogeneity of COST Parameters} \\
For introducing treatment & $Y^{a=1} \independent P \vert Y^{a=0}, V $ 
  \\
For removing treatment &  $Y^{a=0} \independent P \vert Y^{a=1}, V $ 
  \\
  \midrule
\rowgroup{Effect Homogeneity in Distribution} \\
S-ignorability & $Y^{a} \independent P \vert V \; \;  (\forall{a})$ \\
S-admissibility & $Y^{a} \independent P^a \vert V^a \; \; (\forall{a})$  \\
\bottomrule
\end{tabular}
\end{table}
\begin{landscape}
\begin{table}
\begin{center}
\begin{threeparttable}
\tiny
\def~{\hphantom{0}}
\caption{Five Approaches to Effect Transportation}{%
\begin{tabular}
{p{0.35\linewidth}p{0.28\linewidth}p{0.28\linewidth}}
\toprule
&\textit{Interpretation of result} &\textit{Validity Conditions}\\[5pt]
\midrule
 \[\text{RR}_t = \] \[\sum_v \text{RR}_{s,v} \times \text{Pr}(V=v \vert Y^{a=0}=1, P=t)\] &Effect measure in target population. In this specific example, the weights ($\text{Pr}(V=v \vert Y^{a=0}=1, P=t)$) are specific to the risk ratio; similar weights exist for other collapsible effect measures but not for non-collapsible effect measures.
 & Conditional effect homogeneity in measure. (i.e. $V$ must comprise a sufficient set of effect measure modifiers)  \\ 
\midrule
\[\text{Pr} (Y^{a=1}=1\vert P=t)=\]
\[\sum_{v} \left[ \text{Pr}(Y^{a=0}=1\vert P=t,V=v)\times RR_{s, v} \times \text{Pr}(V=v\vert P=t) \right] \] & Average outcome under treatment in target population & Conditional effect homogeneity in measure \\
 \midrule
 \[\text{Pr}(Y^{a}=1 \vert P=t)=\]     
 \[ \sum_{v} \left[\text{Pr}(Y^{a}=1\vert V=v, P=s)\times \text{Pr}(V=v|P=t)\right ]\] &Average outcome under treatment (or under no treatment) in target population & Conditional effect homogeneity in distribution (i.e. $V$ must comprise a set of variables sufficient to block all paths between $P$ and $Y$)\\
  \midrule
   \[\text{Pr}(Y^{a}=1)= \]
  \[ \sum_{v}\left[\frac{\text{Pr}(Y=1 \vert A=a, V=v, P=s) \times \text{Pr}(A=a, V=v, P=s)  }{\text{Pr}(A=a \vert P=s, V=v) \times \text{Pr}(P=s \vert V=v)}\right] \] &Average outcome under treatment (or under no treatment) in population from which the (non-random) sample was taken
 & Conditional effect homogeneity in distribution \\
   \midrule
   \[\text{Pr}(Y^{a} = 1 \vert P \neq s)=\]
  \[\sum_{v}\left[\frac{\text{Pr}(Y=1 \vert A=a, V=v, P=s) \times \text{Pr}(A=a, V=v, P=s)  }{\text{Pr}(A=a \vert P=s, V=v) \times \frac{ \text{Pr} (P=s \vert V=v) }{\text{Pr} (P \neq s \vert V=v)} \times \text{Pr}(P \neq s)}\right] \]& Average outcome under treatment (or under no treatment) in those who were eligible to be selected for the study, but weren't. Note that we do not in general assume that $P$ is binary. In the special situation where $P$ is binary (such that all members of a well-defined source population belong either to the study population or the target population), approach 5 estimates the effect in the same subgroup as approach 3 & Conditional effect homogeneity in distribution \\
\bottomrule
\end{tabular}}
  \begin{tablenotes}
      Note that in approaches 1 through 3, the identifying expressions for the effect in the target population are written in terms of counterfactual variables in order to focus on the part of the analysis that is made necessary in order to account for heterogeneity between populations. In practice, it will be necessary to find an identifying expression that is written terms of observable quantities, which will require either marginal or conditional exchangeability (($Y^a \independent A \vert P=s$ or $Y^a \independent A \vert P=s, V$ for $\forall{a}$) in the study population. In approaches 4 and 5, the identifying expressions are written in terms of observable quantities; these expressions rely upon conditional exchangeability for their derivation. For simplicity of notation, we are here assuming that the same set of variables is sufficient to account both for confounding in the study population and for differences between the populations. 
      \newline \newline
      Note also that the expression in approach 4 can be rewritten on the individual level as $\mathbb{E}\left[\frac{Y \times I(A=a, P=s)}{\text{Pr}(A=a \vert V=v_i, P=s)\times \text{Pr} (P=s \vert V=v_i) }\right]$ in order to illustrate that it can be computed from the data by taking a weighted average in a pseudopopulation where all individuals have been weighted by the inverse of their probability of exposure, and their probability of selection, given their covariates $V$
        \end{tablenotes}
\end{threeparttable}
\label{tablelabel}
\end{center}
\end{table}
\end{landscape}

\section{Effect Homogeneity in Measure}

Effect homogeneity in measure occurs whenever the effect in one population (or subgroup) is equal to the effect in another population (or subgroup) in terms of a particular effect measure, such as the RD or the RR. For example, if the RD in the study population (i.e. $\text{RD}_s$) is equal to RD in the target population (i.e. $\text{RD}_t$) we say that there is effect homogeneity on the RD scale.

Many commonly used methods in epidemiology rely on assumptions that are equivalent to conditional effect homogeneity in measure. For example, the Mantel-Haenszel estimator only has a clear population-level interpretation if the conditional OR is equal between all strata of the covariates \cite{greenland_interpretation_1982} (i.e if there is effect homogeneity in measure between groups in the study population). Epidemiologists also often rely on effect homogeneity in measure when they omit interaction terms from regression models. For example, suppose we fit the the following logistic regression model in the study population:
\[
\text{logit Pr}(Y = 1 \vert A,W, P=s) \ = \beta_0 + \beta_1 A + \beta_2 W.
\]
 
 In this model, by omitting a product term \(\beta_3 \times A\times W \), we encode the assumption that the OR of $A$ on $Y$ in the group $W=1$ is equal to the OR in $W=0$, or in other words, that there is effect homogeneity on the OR scale between two subgroups of the study population: $\text{OR}_{s, w=1} = \text{OR}_{s, w=0}$.

If we are willing to assume homogeneity of an effect measure between two groups in the study population in order to justify the absence of a product term, we may be tempted to ask if we could use a similar homogeneity assumption between one group that is in the study population, and another group that is outside of the study population (e.g. $\text{OR}_{s, v} = \text{OR}_{t, v}$ for all values of $v$) in order to justify extrapolation of an effect to the target population. In this paper, if such a homogeneity condition holds on any scale, we say that there is conditional effect homogeneity on that scale, and that $V$ is a sufficient set of effect measure modifiers for the transportation from the study population to the target population.

The overall idea behind this approach is to identify a set of measured covariates $V$ such that, within levels of the covariates, the magnitude of the effect (when measured on that particular scale) is equal between the populations. To illustrate, it is possible that the RR for adverse effects of Codeine differs between Norway and Japan because the two countries have different distributions of variants of CYP2D6 \cite{bernard_interethnic_2006}, a gene associated with drug metabolism, but that on average, the RR associated with the use of the drug is equal between Norwegians and Japanese who have the same variant of the gene. If that is the case, then we have effect measure homogeneity conditional on CYP2D6 variant, and CYP2D6 is a sufficient set of effect modifiers on the RR scale. Of note, a sufficient set of effect measure modifiers may not exist among the measured covariates. 

 An example of the utility of such homogeneity conditions occurs when an investigator attempts to account for heterogeneity between populations by standardizing effect estimates to the distribution of covariates $V$ in the target population. The first two formulas in table 2 can be used to standardize estimates of a causal effect to a target population, if one has measured a sufficient set of effect modifiers. Approach one, which is a weighted average of the effect measure, is valid for collapsible effect measures \cite{huitfeldt_collapsibility_2019}, whereas approach two, which is a weighted average of the predicted stratum-specific average outcome under treatment, is valid for any effect measure. 

A large literature exists on statistical tests for detecting and quantifying any effect heterogeneity in measure between groups in the observed data (for example, between groups in the study population, or between multiple study populations). Examples of this include Cochran's $Q$ test \cite{cochran_comparison_1950} and the $I^2$ statistic \cite{higgins_measuring_2003}. While homogeneity of an effect measure is to some extent an empirical question \cite{iwasaki_generalizability_2006, poole_is_2015, spiegelman_evaluating_2017}, convincing arguments for stability of the effect measure \textit{outside} of the observed data will often require additional, explicit assumptions about the data generating mechanism. Unfortunately, few examples of data generating mechanisms which result in stability of an effect measure exist in the published literature, and in many settings, finding convincing mechanisms may not be feasible. However, in the next subsection, we discuss the COST parameters framework to demonstrate that at least in some settings, such mechanisms can be found.

\subsection{COST parameters}

COST parameters are a new class of effect parameters that were proposed in order to formalize a counterfactual causal model that may result in effect homogeneity in terms of standard observable measures of effect. The COST parameters for introducing treatment are defined as follows:

\[\text{G} = \text{Pr}(Y^{a=1}=1\mid Y^{a=0}=1)\]

\[\text{H}  = \text{Pr}(Y^{a=1}=0\mid Y^{a=0}=0)\]

The COST parameters can be understood as the proportion of cases and non-cases that would not have had the opposite outcome if their exposure status had been altered. In other words, these are the probabilities that the outcome does not ``switch” in response to treatment (see Figure 1). In Huitfeldt et al \cite{huitfeldt_choice_2018}, it was shown that if certain cofactors that determine treatment effect have equal prevalence between two groups, and if the interaction between these cofactors and treatment $A$ operates according to certain simple biological principles, then the COST parameters for introducing treatment are equal between populations, which can mathematically be written as \(Y^{a=1} \bigCI P \vert Y^{a=0} \). If the cofactors instead interact with treatment according to a different biological mechanism, this would instead result in homogeneity of COST parameters for removing treatment (\(Y^{a=0} \bigCI P \vert Y^{a=1} \)).

\begin{figure}
    \centering
      \label{fig:my_label}
\begin{tikzpicture}[>=stealth',
    shorten > = 1pt,
node distance = 3cm and 4cm,
   every node/.style={fill=white,circle},
              every edge/.style={draw=black,very thick},
    el/.style = {inner sep=2pt, align=left, sloped},
every label/.append style = {font=\tiny}
                    ]
                
\begin{scope}[every node/.style={thick,draw=none}]
    \node (q0) at (-2,3) {Alive if untreated};
    \node (q3) at (-2,-2) {Dead if untreated};
	\node (q2) at (6,3) {Alive if treated};
	\node (q1) at (6,-2) {Dead if treated};
 \end{scope}
 
\path[->] 

    (q0)  edge [bend right=10]  node[el,above]  {$H$}         (q2)
    (q0)  edge [bend right=10]  node[el,below]  {$1-H$}       (q1)
    (q3)  edge [bend left=10]  node[el,above]  {$1-G$}       (q2)
    (q3)  edge [bend left=10]  node[el,below]  {$G$}       (q1)
;
\end{tikzpicture}
  \caption{Counterfactual outcome state transition parameters associated with introduction of treatment. 1-G is the probability that someone who would otherwise die survives if treated. 1-H is the probability that someone who would otherwise survive dies if treated.}
  \end{figure}

Thus, by using the condition $Y^{a=1} \bigCI P \vert Y^{a=0}$ to operationalize effect homogeneity, we reframe homogeneity of the ``magnitude of effect" as a matter of equal distribution of the cofactors that determine whether individuals respond to treatment. If the prevalences of those cofactors differ between populations, such effect equality may hold within levels of covariates $V$, in which case, there is conditional equality of the effect of introducing (or removing) treatment. Conditioning on $V$ is then seen as an attempt to account for those variables that are predictors of the prevalence of the potentially unmeasured background cofactors which determine whether an individual ``switches outcome" in response to treatment.  

Much of the intuition behind traditional approaches to choice of effect modifiers translates readily to the COST parameter framework, but with the added advantage that the line of reasoning is specific to the relevant effect measure. To illustrate, suppose we are interested in generalizing findings about the adverse effects of Codeine. The goal is to account for all cofactors that determine whether a patient will respond to treatment, either by conditioning on those cofactors directly, or by finding observable markers for their prevalence. For example, if the CYP2D6 variant partly determines whether patients respond to Codeine, we could either condition on the gene, or condition on ethnicity as a marker for its prevalence (that is, include either the genetic variant or ethnicity in the set $V$). Further, pre-study drug use may be an observable marker for the prevalence of susceptibility, due to depletion of susceptibles \cite{noauthor_guide_2018}. We will therefore measure and control for ethnicity and pre-study drug use, as well as any other covariates that are relevant according to similar criteria.

COST parameters are generally not identified from the data without strong monotonicity assumptions. If the treatment has a positive monotonic effect (meaning that the treatment does not prevent anyone from having the outcome), and if the COST parameters G and H are equal between the study population and the target population conditional on covariates $V$, there will be homogeneity on the SR scale for exposures which increase the incidence of the outcome (SR becomes equivalent to H, whereas G is trivially 1). Analogous discussion applies when considering a situation in which the treatment of interest is negatively monotonic (protective), in which case RR becomes equivalent to G, and H is trivially 1. 

If treatment has monotonic effects, the COST parameters can therefore be used as a ``bridge" between the biological knowledge on the one side, and homogeneity of observable measures of effects on the other side, thereby allowing the investigator to standardize effect measure from a study to a target population using either approach 1 or approach 2 from Table 2. The necessary weights are discussed elsewhere \cite{huitfeldt_collapsibility_2019}. The bias which is associated with the use of COST parameters in the presence of non-monotonicity is small either if non-monotonicity is negligible, or if the baseline risks are similar between the target population and the study population. If non-monotonicity is not negligible and baseline risks differ between the study population and the target population, the bias associated with COST parameters may be substantial; in such settings, the COST parameter approach should not be used.

The COST parameter approach often results in a recommendation to consider effect homogeneity in terms of the RR scale for exposures that reduce incidence, and in terms of the SR scale for exposures that increase the incidence, while keeping the coding of the exposure such that the ``natural state" of exposure has value 0 and the intervention has value 1. Variations of this suggestion have arisen independently a number of times in the previous literature \cite{sheps_shall_1958, deeks_issues_2002, baker_new_2018}. This approach is also consistent with the Cochrane Handbook \cite{higgins_handbook_2011}, which states that "When the study aims to reduce the incidence of an adverse outcome there is empirical evidence that risk ratios of the adverse outcome are more consistent than risk ratios of the non-event" (the handbook does not take a position on what effect measure to use when the study attempts to estimate the increase in incidence of an adverse outcome). When the disease is rare, this approach is closely approximated by the earlier suggestion to consider ``relative benefits and absolute harms" of interventions \cite{glasziou_evidence_1995}.

Finally, we note an important limitation of COST parameters, which is that they have so far only been defined for binary outcomes. Extensions to continuous and time-to-event outcomes have not yet been established. 

\section{Effect Homogeneity in Distribution}

An alternative approach is to operationalize effect homogeneity in terms of the individual counterfactual distributions under treatment and no treatment. Effect homogeneity in distribution between the study population and the target population holds whenever the following two conditions hold simultaneously: (1) If everyone in both populations were untreated, you would observe the same distribution of outcomes in the two populations ($Y^{a=0} \independent P$) and (2) if everyone in both populations were treated, you would observe the same distribution of outcomes in the two populations  ($Y^{a=1} \independent P$). This condition was referred to as "S-ignorability" by Bareinboim and Pearl, and as "exchangeability between populations" by Lesko et al \cite{lesko_generalizing_2017}. VanderWeele \cite{vanderweele_confounding_2012} showed that effect homogeneity in distribution implies homogeneity of all standard effect measures; effect homogeneity in distribution is therefore a stronger assumption than effect homogeneity in measure on standard scales.

In order to illustrate the difference between effect homogeneity in distribution and effect homogeneity in measure, we again consider the logistic regression model discussed in the previous section:

\[
\text{logit Pr} (Y = 1 \vert A,W, P=s) \ = \beta_0 + \beta_1 A + \beta_2 W
\]

This model is restricted to the study population, and omits the product term \(\beta_3 \times A\times W\). We showed that this model is justified under effect homogeneity in measure on the OR scale between groups in the study population. We note that this model could also be justified under effect homogeneity in distribution between the same two groups. However, this modeling approach has an immediate implication: If effect homogeneity in distribution holds and the effect of $A$ on $Y$ is unconfounded conditional on $W$ and $P=s$, then \(\beta_2\) must be equal to zero (see Appendix 2). This makes the model subject to an empirical test: if e.g. the Wald test rejects $\beta_2 =0$ the model is misspecified. While we do not recommend this as a test of the homogeneity assumption, we believe this example illustrates that effect homogeneity in distribution is a very strong concept, and that investigators often have to rely on a weaker form of effect homogeneity.

As with effect homogeneity in measure, effect homogeneity in distribution may hold within levels of a set of covariates $V$. If effect homogeneity in distribution between the study population and the target population holds conditional on $V$, one can use a third standardization formula (approach 3 in Table 2), based on separately standardizing the conditional risk under treatment and the conditional risk under no treatment from the study population, to the distribution of $V$ in the target population. 

Although methods based on assuming conditional homogeneity of the distribution of a counterfactual variable are mathematically elegant and avoid most of the limitations of defining homogeneity with respect to effect measures, they require strong assumptions which go well beyond the conditions that epidemiologists and clinical scientists have traditionally considered necessary for generalizability. Specifically, whereas approaches that are based on conditional effect homogeneity in measure aim only to control for those covariates that are associated with the magnitude of the effect, methods that rely on conditional effect homogeneity in distribution are valid only if they account for every cause of the outcome that differs between the study population and the target population. In other words, approaches based on conditional effect homogeneity in distribution may lead to biased transportability estimates in the presence of unmeasured causes of the outcome whose distributions differ between the study population and the target population. An example of the implications of such bias was shown recently in the closely related context of agent-based models used for extrapolation\cite{murray_comparison_2017}.  

Effect homogeneity in distribution may occasionally be a reasonable assumption if the imbalance in covariates arises due to a fully understood non-random sampling mechanism, for example, if the investigators enroll participants from an enumerated source population, with selection probability determined by measured baseline covariates. However, outside of such stylized examples, it is more challenging to see good justifications for this type of homogeneity assumption.

We note that approaches based upon effect homogeneity in distribution do not make use of possible information contained in the relationship between what happens if the pharmaceutical is taken, and what happens if the pharmaceutical is not taken. To illustrate, consider a situation where we have conducted a randomized controlled trial on the effect of homeopathy vs no treatment on the incidence of cardiovascular disease, and concluded that the effect in the study population is null. Suppose we are interested in predicting the effect in a different target population, but we believe there may be unmeasured causes of cardiovascular disease that differ between the study population and the target population. In such situations, if we operationalize effect homogeneity using a notion of effect homogeneity in distribution, we are likely forced to conclude that we are unable to make predictions for the target population. In contrast, investigators using an approach based on effect homogeneity in measure could potentially be able to clarify plausible conditions under which the null findings can be extrapolated to the target population. 

\subsection{Weighted estimators for generalizability and transportability}

One particular implementation of generalization based on effect homogeneity in distribution originated with work by Stuart and Cole \cite{stuart_use_2001, cole_generalizing_2010}. These methods extend inverse probability based estimators \cite{horvitz_generalization_1952}, which play a key role in previous work on causal modelling \cite{robins_association_1999, robins_marginal_2000} to the setting of external validity. The validity conditions of these methods are equal to those of standardization based methods discussed above. 

Users of these methods often distinguish between ``transportability" (where the analytic goal is to extrapolate the findings to a target population that does not include those in the study population, i.e. a target population that looks like those who were eligible for, but were \textit{not} sampled in the study), and ``generalizability" (where the target population also includes those in the study population). The methods used for each objective differs in that inverse \textit{probability} of selection weights \cite{lesko_generalizing_2017} (approach 4 in table 2) are used for generalizability, whereas inverse \textit{odds} of selection weights\cite{westreich_transportability_2017} (approach 5 in table 2) are used for transportability. 

Stated slightly differently, the choice between weights is determined by whether the target population is similar to the entire source population from which the study participants were selected, or similar to the subset of the source population that was not selected for the study. We believe the first type of target population is more common; in such settings, inverse probability of selection weights should be used. Inverse odds weights may be appropriate if the study participants are sampled for a pilot study to determine whether the intervention will be implemented in those who were eligible to be selected, but weren't.

Inverse probability weighted methods have been applied to generalize the results of trials on the effect of HIV medication \cite{lu_generalizing_2018} and treatments for substance use disorder \cite{susukida_generalizability_2017}. Lesko et al provided a full description of how these methods can be used in practice \cite{lesko_generalizing_2017}. Buchanan et al \cite{buchanan_generalizing_2018} provided results about the statistical properties of inverse probability weighted estimators for external validity. Dahabreh et al \cite{dahabreh_generalizing_2018} discussed estimators based on augmented inverse probability weights, which are doubly roubst. Nguyen et al \cite{nguyen_sensitivity_2017} showed how to conduct sensitivity analyses on deviations from conditional effect homogeneity. Breskin et al \cite{breskin_using_2019} provided results on bounds, i.e. intervals that show how wrong the point estimates can be in either direction if the assumption of conditional effect homogeneity in distribution is not fully met, in the presence and/or absence of confounding. If one suspects both confounding and lack of effect homogeneity in distribution, these bounds can be used to reason about target validity \cite{westreich_target_2019}, that is, how much bias there may be in the estimates for the target population as a result of deviations both from internal and external validity. 

We note that for collapsible effect measures, re-weighting methods based upon conditional effect homogeneity in measure may be feasible, but to the best of our knowledge, the theory for such methods has not yet been fully developed.
  
\subsection{Causal diagrams for transportability}

One example of a class of data generating mechanisms that guarantees effect homogeneity in distribution (and therefore also effect homogeneity in measure for all standard effect measures) was provided by Bareinboim and Pearl \cite{pearl_transportability_2011, bareinboim_general_2013, pearl_external_2014, bareinboim_causal_2016}, based on causal diagrams \cite{pearl_causality:_2009}. These diagrams are, to our knowledge, the first published formal framework for reasoning about which variables to adjust for when using approaches based on effect homogeneity in distribution. In particular, they use a generalization of effect homogeneity in distribution that allows the covariates that are adjusted for, and membership in the populations that the counterfactual distributions are equal between, to be downstream consequences of treatment.

A selection diagram is constructed as follows: First, the investigator must provide a causal directed acyclic graph (DAG) that is valid both for the study population and for the target population. For this to be possible, the variables must be in the same temporal order between the two populations. If that requirement is met, a DAG which is valid for both populations can be constructed by including every node and edge from the causal DAG in each population. After a shared causal DAG has been constructed, one must also add (1) selection variable nodes ($P$) associated with all variables whose assignment mechanism differs between the study population and the target population, and all variables that depend upon background causes whose distribution differs between the study population and the target population (2) all paths between $P$ and $Y$ that one is not able to rule out based on the temporal structure or expert knowledge. Generally, such paths will exist whenever there are causes of the outcome that differ between the populations. Note that when the goal is to account for ``man-made'' differences between the study population and the target population, i.e. those differences which arise due to how the sample was selected, $P$ is a single binary ``sink node'' representing membership in the study population, which has the same interpretation as the $P$ variable that we have considered so far in this paper. An example of a causal diagram used for transportability is shown in Figure 2. 

\begin{figure}

\begin{tikzpicture}
\node (1) {\textit{U}};
\node [right =of 1] (2) {\textit{A}};
\node [right =of 2] (3) {\textit{Y}};
\node [below =of 2] (5) {\textit{V}};
\node [right =of 5] (4) [regular polygon,regular polygon sides=8, draw] {\textit{P}};

\draw[Arrow] (2.east) -- (3.west);
\draw[Arrow] (1) to [out=25, in=160] (3);
\draw[Arrow] (5.east) -- (4.west);
\draw[Arrow] (5.north) -- (3.south);
\draw[Arrow] (1.east) -- (2.west);

\end{tikzpicture}
\caption{In this causal diagram, findings from the study population may be transported to the target population if we have measured sufficient covariates $V$ to block all paths between the selection node $P$, and the outcome $Y$. We have chosen to represent the selection node $P$ with an octagon.}
\end{figure}

Once a selection diagram has been constructed, one can check for transportability of the results by determining whether $Y$ is d-separated from $P$, given some set of measured covariates $V$, in a manipulated graph where all arrows going into $A$ have been deleted. If such d-separation holds in the manipulated selection diagram, there will exist a transport formula which identifies the causal effect in the target population based on a combination of observed quantities in the study population and observed quantities in the target population. If $V$ consists only of baseline covariates, then the transport formula is equal to the standardization formula discussed in section 3.1. 

Note that the independence relation that is queried by this d-separation approach can be written algebraically as \[
Y^a \independent P^a \vert V^a \; \; \forall{a}\] which Bareinboim and Pearl referred to as “S-admissibility”. When $P$ and $V$ are pre-treatment variables, $P^a = P$ and $V^a = V$  so the independence relation can be simplified as 
\[Y^a \independent P \vert V  \; \; \forall{a}\] (or “S-ignorability”). This simplified version is identical to the previously discussed operationalization of conditional effect homogeneity in distribution, which illustrates the equivalence between the graphical approach and approaches based on standardization or inverse probability weights when $V$ and $P$ are pre-treatment.

Thus, while the graphical approach and the inverse probability weighted approach will result in very similar analyses if $P$ and $V$ are pre-treatment (the analyses will be non-parametrically equivalent but may differ in practice as they may be associated with different modelling assumptions on the joint distribution of variables), the graphical model allows a potentially useful generalization to settings where it is necessary to adjust for post-baseline covariates. In practice, we are not aware of any published examples where a convincing argument was made that a causal effect is transportable only by measuring and adjusting for covariates that were causally affected by treatment. 

Other authors have constructed causal diagrams for generalizability in different ways. In particular, Dahabreh et al \cite{dahabreh_generalizing_2019-1} use Single World Intervention Graphs (SWIGs) to examine conditions under which causal parameters can be generalized from a randomized trial to all trial eligible individuals.

\section{Conclusions}

Causal effects may differ between populations, and investigators will often have to standardize their estimates over a set of effect modifiers in order to make the results applicable to clinically relevant populations. Before it is possible to begin reasoning about which covariates must be standardized over, it is necessary to provide a definition of effect homogeneity. Several different approaches have been proposed.

If effect homogeneity is to be operationalized in terms of stability of a measure of effect, the analytic objective is to account for all those covariates that are associated with the magnitude of the effect on the chosen scale. COST parameters have been developed to formalize conditions that result in homogeneity of observable effect measures. This approach requires that the investigators have accounted for all variables that predict treatment response, that only baseline covariates are necessary for this purpose, and that the effect of treatment is monotonic. When these conditions are met, using COST parameters allows investigators to retain much of the underlying intuition behind traditional approaches to effect modification. Future work may be necessary to develop new classes of causal models that result in homogeneity of other effect measures, including effect measures relevant to time-to-event data. 

If instead effect homogeneity is to be operationalized in terms of conditional homogeneity of the distributions of counterfactual variables (such as in methods based on inverse probability weights and causal diagrams), the analytic objective shifts to accounting for all covariates that are associated with the counterfactual outcome and whose distribution differs between populations. This will generally require a much larger set of covariates. Controlling for all the necessary covariates will sometimes be feasible in situations where the goal is to recover the effect estimates for the full source population in the presence of a fully understood non-random selection mechanism, but may be less realistic in other settings. If the required conditions are met, methods based on effect homogeneity in distribution have considerable advantages, as they do not rely on parametric assumptions or monotonicity conditions. 

All approaches have considerable limitations, and the choice between them will generally depend on expert beliefs about which assumptions are most likely to be approximately true in the specific scientific context.

\bibliographystyle{unsrt}
\bibliography{references.bib}

\section*{Funding}
The authors received no specific funding for this work. Dr. Stensrud is supported by the Research Council of Norway, grant NFR239956/F20 - Analyzing clinical health registries: Improved software 
and mathematics of identifiability. Dr. Swanson is supported by NWO/ZonMw Veni grant (91617066). Dr. Suzuki is supported by Japan Society for the Promotion of Science (KAKENHI grant numbers JP17K17898, JP15K08776, and JP18K10104) and The
Okayama Medical Foundation. Dr. Huitfeldt was supported by the Effective Altruism Hotel Blackpool during revision of the manuscript.

\section*{Acknowledgements}
The authors are grateful to Dr. Issa Dahabreh and two anonymous reviewers for suggestions that greatly improved the manuscript. Any remaining errors are our own. 

\appendix
\section{Appendix 1}
Proofs of identifying expressions from table 2. We note that these proofs are not new to this paper, and are included here only for completeness:

\subsection{Approach 1}

\begin{equation}
\begin{aligned}
&\sum_v{\left[\text{RR}_{s,v}\times \text{Pr}(V=v \vert Y^{a=0}=1, P=t)\right]}\\
&=\sum_v{\left[\text{RR}_{t,v}\times \text{Pr}(V=v \vert Y^{a=0}=1, P=t)\right]} (\because{\text{RR}_{s,v}=\text{RR}_{t,v})}\\
&=\sum_v{\left[ \frac{\text{Pr}(Y^{a=1}=1 \vert V=v, P=t) \times \text{Pr}(V=v \vert Y^{a=0}=1, P=t)}{\text{Pr}(Y^{a=0}=1 \vert V=v, P=t)} \right]} \\
&=\sum_v{\left[\frac{\text{Pr}(Y^{a=1}=1 \vert V=v, P=t)\times \text{Pr}(Y^{a=0}=1 \vert V=v, P=t)\times \text{Pr}(V=v \vert P=t)}{\text{Pr}(Y^{a=0}=1 \vert V=v, P=t)\times \text{Pr}(Y^{a=0}=1 \vert P=t)}\right]}& \\
&=\sum_v{\left[\frac{\text{Pr}(Y^{a=1}=1\vert V=v, P=t)\times \text{Pr}(V=v \vert P=t)}{\text{Pr}(Y^{a=0}=1 \vert P=t)}\right]}\\
&=\frac{\sum_v{\left[\text{Pr}(Y^{a=1}=1\vert V=v, P=t)\times \text{Pr}(V=v \vert P=t)\right]}}{\text{Pr}(Y^{a=0}=1 \vert P=t)}\\
&=\frac{\text{Pr}(Y^{a=1}=1 \vert P=t)}{\text{Pr}(Y^{a=0}=1 \vert P=t)}\\
&=\text{RR}_t 
\end{aligned}
\end{equation}

\subsection{Approach 2}

\begin{equation}
\begin{aligned}
&\sum_v{ \left[ \text{Pr}(Y^{a=0}=1 \vert V=v, P=t) \times \text{RR}_{s,v} \times \text{Pr}(V=v \vert P=t) \right]}\\
&=\sum_v{ \left[ \text{Pr}(Y^{a=0}=1 \vert V=v, P=t) \times \text{RR}_{t,v} \times \text{Pr}(V=v \vert P=t) \right] (\because{\text{RR}_{s,v}=\text{RR}_{t,v})}}\\
&=\sum_v{ \left[ \text{Pr}(Y^{a=0}=1 \vert V=v, P=t) \times \frac{\text{Pr}(Y^{a=1}=1 \vert V=v, P=t)}{\text{Pr}(Y^{a=0}=1 \vert V=v, P=t)} \times \text{Pr}(V=v \vert P=t) \right]}\\
&=\sum_v{ \left[ \text{Pr}(Y^{a=1}=1 \vert V=v, P=t) \times \text{Pr}(V=v \vert P=t) \right]}\\
&=\text{Pr}(Y^{a=1}=1 \vert P=t) 
\end{aligned}
\end{equation}

\subsection{Approach 3}

\begin{equation}
\begin{aligned}
&\sum_{v} \left[\text{Pr}(Y^{a}=1\vert V=v, P=s)\times \text{Pr}(V=v|P=t)\right]\\
&=\sum_{v} \left[\text{Pr}(Y^{a}=1\vert V=v, P=t)\times \text{Pr}(V=v|P=t)\right] (\because{Y^{a} \independent P \vert V \; \forall{a}}) \\
&=\text{Pr}(Y^{a}=1 \vert P=t)
\end{aligned}
\end{equation}

\subsection{Approach 4}

We are here assuming that $Y$ is a binary variable, the proof generalizes readily to settings with continuous or time-to-event outcomes. In order to simplify the logic, we will further assume that the same set of baseline covariates $V$ is sufficient to control both for confounding for $A$, and for differences between populations. In other words, we will assume conditional exchangeability in the study population ($Y^a \independent A \vert V=v, P=s \; \; \forall{a}$) and conditional effect homogeneity in distribution ($Y^a \independent P \vert V=v \; \; \forall{a}$). Before we begin, it is useful to note that \(\frac{\text{Pr}(A=a, V=v, P=s)}{\text{Pr}(A=a \vert P=s, V=v) \times \text{Pr}(P=s \vert V=v)} = \text{Pr}(V=v) \). This follows from sequential application of the definition of conditional probability.

\begin{equation}
\begin{aligned}
&\sum_{v}\left[\frac{\text{Pr}(Y=1 \vert A=a, V=v, P=s) \times \text{Pr}(A=a, V=v, P=s)  }{\text{Pr}(A=a \vert P=s, V=v) \times \text{Pr}(P=s \vert V=v)}\right] \\
&=\sum_{v} \left[\text{Pr}(Y=1\vert A=a, V=v, P=s)\times \text{Pr}(V=v)\right]  \\
&=\sum_{v} \left[\text{Pr}(Y^{a}=1\vert A=a, V=v, P=s)\times \text{Pr}(V=v)\right]
(\because{\text{Consistency}})\\
&=\sum_{v} \left[\text{Pr}(Y^{a}=1\vert V=v, P=s)\times \text{Pr}(V=v)\right](\because{Y^a \independent A \vert V, P=s \; \; \forall{a}}) \\
&=\sum_{v} \left[\text{Pr}(Y^{a}=1\vert V=v)\times \text{Pr}(V=v)\right](\because{Y^a \independent P \vert V \; \; \forall{a}}) \\
&=\text{Pr}(Y^{a}=1) \\
\\ 
\end{aligned}
\end{equation}

\subsection{Approach 5}

The proof of approach 5 is closely related to that for approach 4.  Westreich et al \cite{westreich_transportability_2017} provide a full proof in the appendix.

\section{Appendix 2}

Here, we prove that if there is effect homogeneity in distribution between the groups $W=1$ and $W=0$, then the parameter \(\beta_2\) must be equal to zero in the regression model

\begin{gather}
\text{logit Pr}  (Y = 1 \vert A,W, P=s)  = \beta_0 + \beta_1 A +  \beta_2 W
\end{gather} 

Note here that we are discussing a regression model fit within the study population, and where the homogeneity assumption is between groups of baseline covariate $W$. In contrast to the rest of the paper, we are therefore using the homogeneity assumption $Y^{a} \independent W \vert P=s  \; \; \forall{a}$ rather than $Y^{a} \independent P \vert V=v  \; \; \forall{a}$.

Additionally, we will make the following assumptions:

\[Y^{a} \independent A \vert W, P=s \; \; \forall{a} (\text{Conditional exchangeability in study population}) \]
\[Y^{a} = Y \text{ if } A=a (\text{Consistency}) \]
\\

By consistency and exchangeability, the model can be rewritten as a structural model:

\begin{gather}
\text{logit Pr}(Y^a = 1 \vert W, P=s) = \beta_0 + \beta_1 a + \beta_2 W
\end{gather}

If $W = 0$, we have:

\begin{gather}
\text{logit Pr}(Y^a = 1 \vert W=0, P=s)  = \beta_0 + \beta_1 a
\end{gather} 

If $W = 1$, we have:

\begin{gather}
\text{logit Pr}(Y^a = 1 \vert W=1, P=s) = \beta_0 + \beta_1 a + \beta_2 
\end{gather} 

By the assumption of effect homogeneity in distribution, we can set these equal:

\[\beta_0 + \beta_1 a  =  \beta_0 + \beta_1 a + \beta_2  \]

Solving this for \(\beta_2\) we get \(\beta_2 = 0\).

\end{document}